# An Exploratory Study on Upper-Level Computing Students' Use of Large Language Models as Tools in a Semester-Long Project

Ben Arie Tanay[1], Lexy Arinze[1], Siddhant S. Joshi[1],
Kirsten A. Davis[1], and James C. Davis[2]

School of Engineering Education, Purdue University[1]
School of Electrical & Computer Engineering, Purdue University[2]

## Abstract

**Background:** Large Language Models (LLMs) have begun to influence software engineering practice since the public release of GitHub's Copilot and OpenAI's ChatGPT in 2022. Tools built on LLM technology could revolutionize the way software engineering is practiced, offering interactive "assistants" that can answer questions and prototype software. It falls to software engineering educators to teach future software engineers how to use such tools well, by incorporating them into their pedagogy.

While some institutions have banned ChatGPT, other institutions have opted to issue guidelines for its use. Additionally, researchers have proposed strategies to address potential issues in the educational and professional use of LLMs. As of yet, there have been few studies that report on the use of LLMs in the classroom. It is, therefore, important to evaluate students' perception of LLMs and possible ways of adapting the computing curriculum to these shifting paradigms.

**Purpose:** The purpose of this study is to explore computing students' experiences and approaches to using LLMs during a semester-long software engineering project. We investigated the impacts of a low-cost intervention. While there have been studies on the use of LLMs in the classroom, there have been limited works on the use within a project-based course in the computing classroom. Our study helps fill this knowledge gap.

**Design/Method:** We collected data from a senior-level software engineering course at Purdue University, a large public R1 university in the Midwest. This course uses a project-based learning (PBL) design with a semester-long team project. In Fall 2023, the students were required to use LLMs such as ChatGPT and Copilot as they completed their projects. A sample of these student teams were interviewed in the middle and at the end of the semester to understand: (1) how they used LLMs in their projects; and (2) whether and how their perspectives on LLMs changed over the course of the semester. We analyzed the data qualitatively to identify themes related to students' usage patterns and learning outcomes.

**Results/Discussion:** We report on students' thinking over the course of the semester and how they developed strategies to use LLMs. Our results characterize the impact that the incorporation of LLMs had on the students' learning. We show that when computing students utilize LLMs within a project, their use cases cover both technical and professional applications. In addition, these students perceive LLMs to be efficient tools in obtaining information and completion of tasks. However, there were concerns about the responsible use of LLMs without being detrimental to their own learning outcomes. Based on our findings, we recommend future research to investigate the usage of LLM's in lower-level computer engineering courses to understand whether and how LLMs can be integrated as a learning aid without hurting the learning outcomes.

## Introduction

In every generation, software engineering education must adapt to technological innovations. In our generation, we must respond to large language models (LLMs). LLMs are machine learning models (typically with billions of parameters) that are trained on vast amounts of data [1]. They are known for their ability to generate human-like text and can be used in a variety of tasks such as code synthesis, conditional text generation, and mathematical reasoning [1], [2]. Due to their strong performance on a variety of tasks, LLMs have found diverse uses in both academia and industry [3], [4]. Notably, OpenAI's ChatGPT and GitHub's Copilot are built on LLMs technology and are widely used by instructors, researchers, and software engineers.

These LLM-based tools have influenced student behavior as well. Students use them in research and writing, as well as study guides and even in lieu of teaching assistants [5]. Particularly in the field of computing, students have found them useful for tasks such as code generation, summarization, and explanation [6], [7]. This mass adoption by university students has prompted a range of opinions and perspectives from various stakeholders, spanning from rejection to acceptance with caution. Researchers have also developed varied recommendations for guidelines and policies aimed at maximizing their utilization [8], [9], [10]. Additionally, studies have been conducted to explore students' perspectives on the use and impacts of LLMs, especially within introductory level courses [11],[12],[13].

This study investigates how upper-level computing students utilize LLMs as tools in a semester-long project. We conducted in-depth discussions with the students, exploring how their experiences evolved over the semester. We specifically investigated how the students used these tools in their project, the strategies they employed, and the impact of this usage on their learning experience. Qualitative analysis was used to address two research questions: RQ1: How do students integrate LLMs into coursework when policies allow unrestricted access? RQ2: How does the use of LLMs influence students' perceptions of their learning?

Our results demonstrate that students use LLMs for an array of different tasks. Students used LLMs for both technical and professional tasks, including programming support, idea generation, writing support, and project management. Students found that LLMs increased their productivity by providing easier access to information and solutions, which allowed students to become more self-sufficient. However, students also discussed potential misuse of LLMs, specifically their concerns about developing a reliance on the technology or not having the appropriate prerequisite knowledge.

# Background & Related Work

## LLM Technology and Applications

Large Language Models (LLMs) are a recent advancement in natural language processing (NLP). These models are trained on large amount of textual data to understand and generate human-like language [7]. While early language models primarily focused on text generation (e.g., advertising copy), LLMs can tackle more complex tasks [1]. One application of interest is in interacting with computer programs, *i.e.*, software, which can be represented with specialized text [14]. While a wide array of LLM applications exist today, this research only focuses on two: ChatGPT and GitHub Copilot. A brief description of these tools can be found in Figures 1 and 2.

**Figure 1**
*ChatGPT [15], a general chatbot developed by OpenAI. Capable of sustaining conversations with the user.*

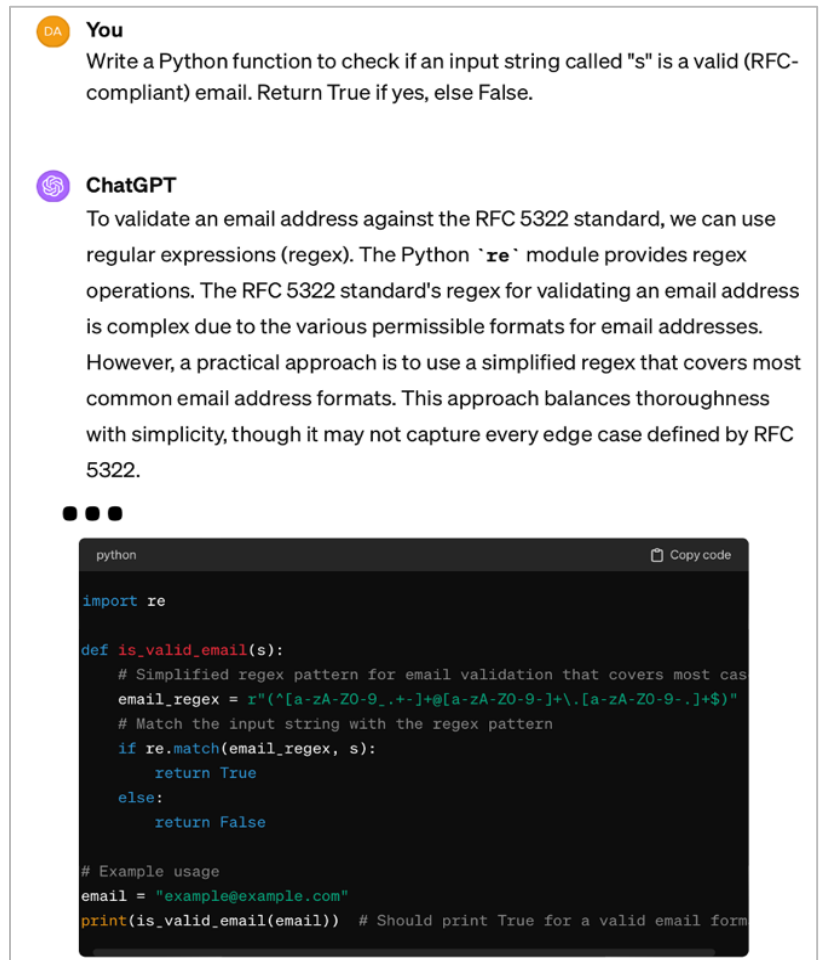

*In this example, we provided the prompt shown at the beginning, and ChatGPT v4 replied. Two excerpts are shown. Note that ChatGPT v4 correctly identifies the RFC for emails, RFC 5322.*

**Figure 2**
*GitHub Copilot [14]. A code autocompletion tool developed by Microsoft and OpenAI. Integrates into popular IDEs (e.g., Visual Studio Code) from installation. Capable of generating code based on comments and previous code written by the user.*

```python
# Check if an input string called "s" is a valid
  (RFC-compliant) email. Return True if yes, else False
# Example: s = "a@gmail.com" -> True
# s = "x@xxxxx" -> False
# s = "x@x.x" -> True
# s = "x@x.x.x" -> True

import re

def is_email(s):
    # regex pattern for email
    pattern = r"^[a-zA-Z0-9._%+-]+@[a-zA-Z0-9.-]+\.[a-zA-Z]{2,}$"
    return bool(re.match(pattern, s))
```

*In this example, we provided Copilot with the prompt indicated on line 1. The remaining lines were generated entirely by Copilot via a succession of auto-complete suggestions. The body of the is_email function is grey and opaque, signifying a suggestion that the engineer is considering.*

LLMs are being adopted in both industry and academia owing to their performance on various software engineering tasks, such as requirements elicitation and refinement, specification, implementing a design, and validation [16], [17]. In industry, they are tools that can be used to enhance productivity [18]. In computing education, instructors have found applications for LLMs as educational aids [19], generators for programming assignments, and tools for providing code explanations [20]. Meanwhile, students have found value in utilizing LLMs for research and academic writing, as well as for idea and code generation [5].

Similar to any technological advancement, there are educational challenges inherent in the use of LLMs, including students' overreliance, issues of plagiarism, and biases in the generated content [21]. These challenges highlight the need for policies and guidelines towards the responsible use of LLMs. Educational institutions are actively engaged in deliberations to determine the most effective strategies for incorporating LLMs into their curricula [8]. As institutions grapple with this decision, there have been noteworthy efforts to provide guidance on the ethical use of LLMs [8],[22],[9],[23]. Researchers caution against implementing "one-size-fits-all" policies but advocate for the adoption of flexible policies tailored to specific contexts, applications, and disciplines [10],[24]. In the present study's institution, there is no university policy on the use of LLMs [25]. Instructors are given autonomy to set policies, and the institution provides guidance and example syllabi to assist them.

**LLMs in the Classroom**

LLMs have found diverse applications in the classroom. Some instructors have employed LLMs to create quizzes and flashcards, aiming to improve student learning and assist with exam preparation [26], [27]. Ngo [12] conducted surveys and interviews to explore students' perspectives on using ChatGPT for learning and found that students generally had a positive opinion of the tool. Haensch et al. [28] obtained similar results by analyzing TikTok videos related to ChatGPT in February 2023 and found that majority of the videos had a positive outlook on ChatGPT and had the most likes by users. In a recent study, Kung et al. evaluated the performance of ChatGPT on the United States Medical Licensing Exam (USMLE), identifying its potential to contribute to medical education and clinical decision-making. These studies are indicative of the versatility and impact of LLMs across various fields of education.

In software engineering classrooms, researchers have examined various applications of LLMs. Jalil et al. used prompts from five chapters of a popular software testing textbook to demonstrate how ChatGPT could serve as a valuable guide for students [29]. Davis et al. performed a similar study with questions suited for introductory programming courses in C [30]. Other researchers have explored students' use of LLMs and their varied perceptions. Liu et al. integrated AI tools in an introductory course (Harvard CS50) to aid teaching and learning [31]. A user study by Vaithilingam et al. explored how students and programmers utilize and perceive Co-pilot [32]. Considering that students use Copilot to learn code, Puryear and Sprint investigated its impact on students' code learning process within introductory computer science and data science courses [11]. In another online introductory programming course, Hellas et al. assessed the effectiveness of LLMs in identifying issues within the code on which students commonly seek help [33]. Liffiton et al. have implemented an LLM named CodeHelp in a first-year computer and data science course over a 12-week period to understand how students' LLMs perceptions and usage patterns change over time [13].

In summary, prior studies on how students use LLMs in the classroom have focused on introductory courses when students are first learning programming languages. There has been less research on upper-level undergraduate students, particularly in a project-based classroom setting. Given the various roles that LLMs can play in software engineering projects [34], [16], [35], we wanted to understand how undergraduates studying software engineering approach the use of LLMs in their projects and their perceptions of how LLMs impact the overall project outcomes. Our study contributes insights into LLM use at the stage when students are moving beyond learning programming to applying these skills in a software engineering project.

## Course Overview

**Course Structure & Project**

"ECE 461: Software Engineering" is a senior-level course for electrical and computer engineering students at Purdue University [36]. The prerequisite coursework is two courses in programming and one course in data structures and algorithms, all taught in C. Most students have also taken a course in Python programming. The course learning outcomes are (1) an understanding of common models of the software engineering process (e.g., agile methods, plan-based methods); (2) the ability to conduct the software engineering process (e.g., requirements elicitation, project specification, design, implementation, validation, maintenance and evolution, re-use, and security analysis); and (3) an understanding of the social aspects of software engineering, (e.g., teamwork and ethics).

The course uses project-based learning [37] to teach these learning outcomes. Students work on a course project in small teams in teams of four that spans the entire 16-week semester[1] . The project has two phases. Phase 1 takes 4 weeks. After Phase 1, the teams exchange projects (to simulate brownfield engineering [38]) and undertake the 12-week Phase 2 on top of another team's implementation of Phase 1. During Phase 1, the students are applying concepts they have learned in previous coursework, such as file I/O and command-line interfaces. In Phase 2, the course staff provide support as students self-learn modern software engineering practices (e.g., continuous integration and continuous deployment) and cloud computing technologies (e.g., component selection and integration via infrastructure-as-code). The majority of the implementation is required to be done in TypeScript, a programming language that is not covered in the curriculum.

**LLM Policy**

In light of the advent of LLMs, the course staff revised the syllabus in the Fall 2023 offering of ECE 461. The following snippet from the Fall 2023 syllabus outlines the course policy regarding the use of LLM tools.

> *"Despite their limitations, [LLM] tools are already transforming the discipline of software engineering. Engineers who figure out how to use them will get promoted. Those who do not will miss out on opportunities. Therefore, the use of these tools is mandatory…You will be required to use such tools as part of your project, in a manner that you and your team will determine."*

In addition to the syllabus, the project assignment reaffirmed the requirement to use LLMs and provided examples for students to explore.

> *"Your team must use a large language model. I recommend GitHub's Copilot or Meta's Code Llama. Your Project Plan should include a description of how you used the LLM in a responsible way."*

Although the LLM policy allows students to use LLMs in whatever capacity they see necessary, the course syllabus includes a policy on plagiarism that (deliberately) does not mention LLMs.

---

[1] Relevant course materials, such as the syllabus, project description, and prompt engineering materials mentioned in the LLM Pedagogy section, are available at https://davisjam.github.io/teaching.

The following snippet defines plagiarism and the course instructor's guidelines on how to ethically re-use an original work:

> "In each kind of assignment, the primary consideration is that you clearly indicate which parts of your submission are your own work, and which parts are communicating someone else's work. A failure to make this distinction is commonly called plagiarism. However, in the engineering workplace, what academics call 'plagiarism' is usually thought of as 'benefitting from someone else's expertise'. Engineering knowledge is communal expertise hard-won over many years. With this in mind, I am open – indeed desirous – to see you learn how to re-use concepts and code. But thoughtfully! In your assignments, you must justify your decisions. This includes re-use decisions, e.g. of designs, of components, or of tests."

**LLM Pedagogy**

To support students in using LLMs, the course staff developed a learning module that discussed LLMs and their potential use in software engineering activities in general and in the course project in particular. Guided by works such as [17], the lecture described LLM technology and ways to interact with an LLM to refine requirements, specify a system, and simulate it. The accompanying homework had students apply these concepts to develop and begin implementing a test plan for their project using the ChatGPT LLM. These pedagogical materials will accompany the final version of the paper.

## Research Questions

In this study, we sought to understand how students in an upper-level software engineering course interacted with LLMs in their academic work. More specifically, we studied two points of interest: student-developed use cases for LLM operations and student-perceived impacts that LLMs have on their learning. Therefore, we investigated two research questions:

RQ1: How do students integrate LLMs into coursework when policies allow unrestricted access?

RQ2: How does the use of LLMs influence students' perceptions of their learning?

## Methods

We conducted interviews with students in the ECE 461 course at two points in the Fall 2023 semester: once at the halfway point and once at the end. The interview transcripts were then thematically reviewed in a multi-step process, and then analyzed and interpreted. In this section we describe the participants, data collection, and data analysis.

**Participant Recruitment**

At the midway point of the Fall 2023 semester, students were briefed on the study during an in-class presentation by one of the research group members. The presentation covered the researchers' motivations for the study, what participation would entail for those who opted into the study, and the financial incentives for participation. In line with our IRB protocol (Purdue

#2023-1460), students were assured that their participation in the study was entirely optional and that their identities would remain anonymous to their course instructor until the end of the semester. This process was followed to ensure that class grades would not be impacted by students' decisions to engage with the study or not. During the in-class presentation, a survey link was provided where students could sign up to participate in the study. Students who signed up via the survey were then contacted by members of the research team to schedule interviews. Participants were given a $20 gift card for each interview that they completed. Of the 72 students enrolled, nine registered for interviews. Of the 17 total teams, five were represented by one participant and two teams were represented by two participants. Demographic information for the course and the participants can be found in Table 1.

**Table 1**
*Participant and class-wide demographics.*

| Major | Participants | Class-wide |
|---|---|---|
| Computer engineering | 9 | 75 |
| Other | 0 | 0 |
| Class standing | | |
| Senior | 7 | 65 |
| Junior | 2 | 10 |
| Other | 0 | 0 |
| Gender | | |
| Man | 9 | 69 |
| Woman | 0 | 6 |
| Other | 0 | 0 |

**Data Collection**

In this subsection we discuss our interview protocol development process followed by the steps we took to collect longitudinal interview data from the participants. The process of interview protocol development and its refinement is crucial to enhance the quality of data collected in a study [39]. Longitudinal interviews help identify changes over time, explore how the change occurred, and perspectives of the individual who experienced the change [40].

*Interview Protocol Development*

To develop the interview protocol, we first identified the main constructs that were of interest to our study. These main constructs were developed based on our literature review of similar previous studies [33], gaps identified from our literature review, and the ECE 461 instructor's experience on teaching LLM usage in the classroom (they are also a co-author of this work). As such, we decided to focus on three constructs in our interviews (1) past experiences with using LLM's before entering the course; (2) experience using LLM's in the ECE 461 course project; and (3) Interpretation of experiences using LLM's in coursework or projects. Once we identified these constructs, we developed an initial version of our protocol by iteratively discussing the interview questions and receiving feedback from senior members of our research team. Subsequently, one member of the research team conducted a pilot interview with the graduate teaching assistant of the ECE 461 course. This student had previously taken the ECE 461 course. After the pilot interview, the research team reconvened to refine the protocol and determine if the protocol was targeted to answer the research questions. For example, we rephrased questions to eliminate the possibility of one-word answers (e.g, "yes" or "no"). We also added questions to gain more context on students' relationships with LLMs before enrolling in ECE 461.

*Participant Interviews*

We conducted two rounds of in-person semi-structured interviews, with the same participants in both rounds. All participants consented to participate in the study. We conducted the first round of interviews during weeks 9-10 of the Fall 2023 semester, and the second round of interviews during weeks 15-16 of the same semester. Two researchers collaborated to conduct the interviews; each interview was conducted by one of them. The interviews for our study typically lasted for an hour and were audio recorded.

As noted above, the interview protocol focused on three constructs: (1) participants' experience with LLMs prior to ECE 461; (2) their experiences in using LLMs in ECE 461 coursework; and (3) their interpretations of how LLMs influenced their coursework and learning. Because we used a two-phase interview design, we varied the interview protocol slightly between the two rounds. The second round of interviews focused more on use cases of LLM's and experiential interpretations, and less on past experiences. Additionally, we asked the participants to compare their experience with the LLMs between the first and second phase of the course project because the nature of the software engineering tasks changed between the two phases. Table 2 provides examples of the types of questions asked in each round of interviews. Both rounds' interview protocols will be provided in the appendix.

**Table 2**
*Example Interview questions.*

| Round | Set | Number of Questions | Example |
|---|---|---|---|
| 1 (Halfway into semester) | Past Experiences | 6 | "How did you first find out about LLMs? What made you want to explore LLM tools?" |
| | Use Cases | 6 | "Tell us about your typical use of LLMs in this project, and the kinds of value you get from them." |
| | Interpretations | 3 | "Do you feel that the use of LLMs will enhance your learning experience? Why or why not?" |
| 2 (Last 2 weeks of semester) | Past Experiences | 2 | "What are your current impressions of LLMs as compared to when you first started using them? |

| | Use Cases | 11 | "Was there a substantial difference in how you used LLMs between phases in the project?" |
| | Interpretations | 5 | "How do you see the skills of knowledge you gained from using LLMs benefitting your future coursework or projects?" |

**Data Analysis**

We had the interviews transcribed using an online transcription service called Rev.com. Upon transcription, the research team vetted the transcripts by repeatedly reading them and checking them against the audio recording. We fixed any discrepancies between the transcripts and audio recordings before proceeding with the data analysis.

To address our research questions, we used a thematic analysis approach. Thematic analysis is a data analysis technique that is driven by the research questions of the study and aims to identify and report themes emerging from the qualitative data [41]. We first divided the transcripts between two researchers who conducted the interviews. To promote full knowledge of the data, each researcher reviewed the transcripts of the interviews conducted by the other researcher. While reviewing the transcripts, the researchers documented memos of key ideas, common responses, and meaningful quotes. In discussion with the larger research team, these memos were grouped into distinctive buckets and a set of initial codes were generated from the data set. The two researchers who coded the data then met to discuss the codes and come to a consensus on the coding scheme. These researchers then coded the full set of transcripts using the agreed upon codes. After reviewing the quotes aligned with each code, the two coders identified potential themes in response to each research question. These potential themes were discussed by the whole research team to generate a final set of themes and subthemes.

## Results

We present the results of each research question separately with the main themes we developed in response to that question. In the first section, we discuss how students used LLMs in their course project (RQ1) and in the next how they perceived that LLMs impacted their learning processes (RQ2).

**RQ1: How do students integrate LLMs into coursework when policies allow unrestricted access?**

Our first question explored how students used LLMs in their projects. From our thematic analysis, we identified two unique themes in their responses, each with two subthemes. We summarized these themes and the number of students who mentioned each theme in Table 3 below.

**Table 3**
*Uses of LLMs described in student interviews.*

| Theme | Subtheme | Definition | Frequency |
|---|---|---|---|
| Technical aid | Programming support | Student used LLMs for technical assistance in writing/editing code, understanding new software languages, and learning new software engineering concepts | 9 |
| | Idea generation | Student used LLMs for creative tasks such as designing a system, understanding/following best practices, and approaching complex problems | 5 |
| Professional aid | Writing support | Student used LLMs for communication assistance in emailing instructional faculty and writing assignments | 6 |
| | Project management | Student used LLMs for organizational assistance in planning division of labor and timelines for project milestones | 3 |

*1. Programming support*

All nine students used LLMs to help them generate code or modify their pre-existing solutions. In the following example, Participant 9 provides an example of how he interacted with ChatGPT to generate code in TypeScript, a language he had no prior experience with:

> *"So I just told it, 'Okay, show me TypeScript code to find out if...' Actually, first I asked it how I could do it, how are versions of dependencies measured, and how to find out if they were constrained or not constrained. And then once I had learned enough about it, and considering the fact that I already knew how to fetch data from GitHub API, I just straight up asked it, 'Give me TypeScript code which fetches whether dependencies are constrained or non-constrained,' based on the earlier interactions that I already did. And... Yes. It showed me code for that."* [Par. 9]

In this case, Par. 9 demonstrates a common interaction protocol when interacting with ChatGPT: first, provide the LLM with context for the problem, then request a technical solution by asking a specific prompt. This simple approach to LLM interactions was first introduced to students by the course instructor as a part of the course materials. In the following example, another student explained how they used GitHub Copilot in a similar manner:

> *"[I used] a little bit of ChatGPT for understanding the framework of TypeScript, getting started with TypeScript. And there was a lot of Copilot for, 'I need to write this function.' I write a comment for this function, see what it gives me." [Par. 3]*

Par. 3 notes that he used ChatGPT and Copilot for fundamentally different roles in the project, which was a sentiment echoed by other students as well. While students restricted the use of Copilot to simply act as a code generator and editor, they regularly assigned ChatGPT to other complete other complex tasks beyond coding, as exemplified below. These tendencies align with the descriptions of these two LLM tools in Figures 1 and 2.

### *2. Writing support*

Six students used LLMs to benefit their writing and communication endeavors. This example highlights how ChatGPT was used in such a manner:

> *"So I think [ChatGPT] helped me format my emails, help me keep a professional tone. So now it's a lot easier to write an email than it was a year ago. Before a year ago, I'd have to be like, 'Oh, shoot. What word do I want to use? How do I start off a sentence?' But because I've asked ChatGPT to help me, I know the general structure that it will generate." [Par. 8]*

Par. 8 expressed in this snippet the desire to establish a professional tone adequate for emails. Most commonly, students cited using LLMs to edit similar correspondences, such as generate emails to the professor or check for grammatical errors on an assignment.

### *3. Idea generation*

Five students used LLMs to play a creative role throughout their projects. In this example, one of the students shared his process for using ChatGPT to help him construct an alternative solution to the one he presented:

> *"And it could also serve a rubber ducky[2] kind of role where I'm just telling it, 'Okay, this is what I'm trying, this is what is not working.' And even if it doesn't really know what next steps or what's the exact line of code I need to include, it can, I guess, I don't want to say reason just because I know that an LLM cannot reason, but it can provide alternative ways that I can use, which I've not thought about." [Par. 6]*

Par. 6 demonstrates one of the ways he interfaces with LLMs. In this case, he provides the LLM with an appropriate amount of context of both his problem, then requests guidance towards a solution rather than prompting for a specific, technical fix. Other students demonstrated the use of LLMs for directional decisions, like so:

> *"I'm much more detail-oriented [...] I use [ChatGPT] to get a broader picture, and [I] try to focus on components. And then, when I'm actually trying to get something integrated and working with other aspects, I am more detail-oriented there." [Par. 1]*

---

[2] Rubber ducky is a debugging technique in programming where programmers use a rubber duck or an inanimate object as a sort of listener as they explain their codes line by line. This helps them vocalize the logic behind their codes and identify bugs.

Par. 1 shared his preferences for writing his own code but chose to outsource the scoping of the project to LLMs. Of all the students who used LLMs for idea generation, their use cases could fit into one of two buckets: idea generation for system-level design, or idea generation for debugging and editing their solutions.

*4. Project management*

Three students indicated that they used LLMs to assist in administrative duties, such as project management and organizational planning. The following quote demonstrates how one of the students tasked ChatGPT with generating a timeline and the division for labor of the project at the start of the semester:

> *"I told ChatGPT to describe, 'You are a project manager. Propose a timeline and split the work to four people.' And it correctly identified a front end, a backend, a DevOps, and I believe it was a AWS/security person, and then, it split the work according to that." [Par. 6]*

Par. 6 gave exact instruction to ChatGPT for generating a project timeline from the perspective of a manager and was interviewee to report doing so. However, other students discussed the role of LLMs for other tasks related to professional skills like organization:

> *"[ChatGPT] was more of an organization tool. And so, it saved a lot of time for me because, instead of spending an hour [planning], I just had it laid out in front of me." [Par. 2]*

While the theme of project management was the least common among the four use cases identified, all three students had unique ways of demonstrating it. Par. 6 notably stood out from the other two participants due to how he assigned an identity to ChatGPT in saying "you are a project manager." This phenomenon demonstrated itself in the speech of other students throughout the interviews and will be discussed more thoroughly in the Discussion section.

**RQ 2: How does the use of LLMs influence students' perceptions of their learning?**

Our second question explored how students perceived LLMs influencing their learning from the thematic analysis, we identified six unique themes within their responses: knowledge retention concerns, over-reliance on LLMs, improve accessibility of information, improved accessibility of solution, requires prerequisite knowledge, and improved self-sufficiency. We summarize these themes and how frequently they were mentioned by interviewees in Table 4 below.

**Table 4**
*Themes describing students' perceptions of how LLMs impact their learning.*

| Theme | Definition | Frequency |
|---|---|---|
| Accessibility of information | Student believes that using LLMs allows them to research and gather intelligence more efficiently | 8 |
| Accessibility of solution | Student believes that using LLMs allows them to design successful project elements faster | 8 |
| Knowledge retention concerns | Student believes that using LLMs may hurt their ability to retain novel concepts. | 8 |
| Requires pre-requisite knowledge | Student believes that using LLMs may be ineffective if users do not have a base-level understanding of certain topics | 6 |
| Increased self-sufficiency | Student believes that using LLMs allowed them to decrease their reliance on help from course staff | 4 |
| Over-reliance on LLMs | Student believes that using LLMs may create a dependency on the technology to find solutions | 3 |

*1. Improved accessibility of information*

Eight students cited that using LLMs greatly facilitated the knowledge acquisition process. One of these students made note of his abilities to retain new LLM-provided information in the provided snippet:

> *"So, I think it's one of those things where I learn this knowledge and then I just kind of keep it with me. So, I think a big thing that it might help with is, like any machine learning AI, jobs and interviews. I now know this topic and, even though you can say, 'Oh, you didn't go through as much reading and textbooks. You just read one paragraph and you're going to forget it.' But I think that the majority of learning, especially for me, comes from kind of using it and exploring with it." [Par. 2]*

Here, Par. 2 is aware of how much nuance LLMs omit when they generate content. He then goes on to explain that being exposed to the new knowledge in the first place and interacting with it via prompt engineering is just as valuable to his learning. Other students felt a similar level of satisfaction with LLMs, as demonstrated here:

> *"[…] as a research tool, it's so powerful and you can easily find a lot of information at your fingertips using ChatGPT and focused information rather than general information. So much easier than if you use the web. Because the web you kind of have to go through and find exactly what you need, but with this you can easily find what you need." [Par. 7]*

Just like Par. 2, Par. 7 uses LLMs like ChatGPT as an effective way to consolidate information from multiple sources to a single webpage. Par. 7 is also particularly satisfied with the level of depth and complexity that ChatGPT responds to his queries with. Many other participants' responses cited the accessible nature of ChatGPT for why they often preferred interfacing with it instead of traditional search engines like Google.

*2. Improved accessibility of solution*

Eight students also cited that LLMs played a key role in identifying and implementing software engineering solutions to their projects. The following quote provides insight on how LLMs can be used to reduce the amount of time to find solutions:

> *"It's more just that using ChatGPT, everything is in one place and I have quick access to asking a question. It gives me the response, as opposed to spending five minutes hunting down the official documentation or example usages in a YouTube video or Stack Overflow or something like that." [Par. 1]*

In this case, Par. 1 demonstrates that ChatGPT can be faster information aggregators than traditional community forums. This in turn allows him to implement a solution more quickly than before, especially without having to "hunt down" official documentation. In the following example, another student recounts similar levels of increased productivity when working with Copilot to generate text:

> *"[I'm] getting two to three times as much done [by using Copilot] as I would [have by] writing every line. [...] I typed the comment and then in five seconds I have 10 lines of function versus that's going to take me two minutes to write 10 lines of function." [Par. 3]*

Par. 3's comment is representative of a general sentiment among students that that they solved problems faster when they included LLMs into their workflow.

*3. Knowledge retention concerns*

Eight students acknowledged that they were unsure if their LLM usage contributed to a lack of knowledge retention. Throughout the interviews, students revealed that at times they interacted with LLM solely with the intention of retrieving a solution to implement into their project without genuinely learning the content. For example, one student explained:

> *"It's definitely damaged my learning experience, in some cases, where if I'm doing the idea generation portion, I do actually learn things like this is good, this is bad for this scenario or whatever. But if I'm just doing the technical side of things, trying to complete a task that I've already planned out, I don't really retain those methods as well. I'm just trying to make it work." [Par. 1]*

In this example, Par. 1 acknowledges that his retention of knowledge varies based on the task at hand. When using LLMs to help with the creative process, the participant believed this to be an academically enriching experience. However, in technical applications such as implementing the code outlined from the creative/planning process, the participant's behavior suggests their belief that the solution can be achieved without the need for understanding the technical concepts. Another student shares similar thoughts with more explicit mention of his concern:

> *"I feel like I'm writing code and Copilot suggests something and I'm looking at it, I'm like, 'Yep, that looks good and I'll accept it' [...] I make sure that I'm reading the code before I* *accept it, but it's also like, 'Well now I don't even have to think about what I'm doing because it just knows what I'm going to do.' So I don't know. It's a little concerning." [Par. 5]*

Par. 5 notes that although he does review the code generated by LLMs, he suggests that the review process is more optional now that he "[doesn't] even have to think about" the code he's implementing. In fact, other participants openly discussed this issue, for example:

*"I think I'm relying on it maybe a little bit too much in [ECE 461, because I'm not trying to really understand everything about what code is producing, and I'm like, "Okay, that worked." And then I always ... I'll copy it, and then I'll just go back and understand it, and I never do." [Par. 4]*

Par. 4 is not alone in admitting to his tendency to deprioritize learning the content generated by his queries, but this example further demonstrates that all of the participants share a belief that using LLMs has, at minimum, the capability to influence knowledge retention. Whether these retention challenges stem from the users or the LLMs is where many participants' opinions become more nuanced. These nuances in participants' opinions are further discussed under subsequent themes.

*4. Requires prerequisite knowledge*

Six students said that extracting the most value out of LLMs requires a minimum amount of background experience or information relevant to the project. When asked where in their department's curriculum should students be exposed to LLMs, one student stated the following:

*"I don't think [LLMs] should be applied to a class like [data structures and algorithms class, a prerequisite for this course], because even though that's a high-level programming class, the purpose of [it] is to learn C and learn the algorithms. But knowing the algorithms and not knowing the C to write the algorithms, does no good. Whereas [with ECE 461], you've already built that foundation [of knowing how to write C]. Now it's applying it to how we will use in the workplace." [Par. 3]*

Par. 3's point is that learning algorithms from LLMs without having a foundational working knowledge of the C language can be ineffective. His concern for the mis-adoption of LLMs does not extend to his experience in ECE 461 however, as he believed that the foundations required for ECE 461 have already been built by previous classes. Other participants shared similar concerns towards introducing LLMs to a student before a solid foundation of knowledge can form. For example:

*"But I do think that students, if they're learning how to code, they should learn the fundamentals first before learning how to use [Chat]GPT. [...] [LLMs could be introduced to students in] maybe middle school or high school, some point where they can recognize how powerful the tool is, and how it can be good and how it can be really bad." [Par. 7]*

While Par. 7's initially refers to prerequisite technical knowledge like the ability to understand code, he also makes note of prerequisite conceptual knowledge, like the full extent of what ChatGPT and other LLMs are capable of. Par. 7 also warns that LLMs can be used in a way described as "really bad," suggesting his belief that misuse of LLMs can be damaging as well. He then provided an example of a elementary school student who would use LLMs to get their homework done quickly in order to spend more time on recreation.

### *5. Improved self-sufficiency*

Four students noted that their LLM usage resulted in them needing to seek help from instructional staff less than expected. One student explained how much more accessible LLM technologies can be compared to their human counterparts in the classroom environment:

> *"Teaching assistants require resources that are very limited at times. Or maybe you're working on [an] assignment at a time when teaching assistants aren't available, Piazza is not active. So, you're using [ChatGPT] as another resource. There's Stack Overflow. It's another resource like that. And I've used it this way to where it's very helpful to get another explanation, or, 'I'm struggling with this. Give me an example for this topic and how to solve it.'" [Par. 3]*

Par. 3 notes that LLMs are another resource where a student's access is not dependent on the schedules and time of others. This perspective suggests that the accessibility of LLMs grants students the opportunity to have more sovereignty over the time they allot themselves to work on their projects. While previous results already demonstrated that students were saving more time by using LLMs instead of search engines or traditional online research, another student remarked that LLMs can be another time-conscientious alternative to course staff:

> *"[…] I'm getting explanations for the things I didn't really understand a lot quicker than I would just kind of looking it up, or going to a TA for help or the professor" [Par. 2]*

Par. 2 then goes on to provide additional remarks where he makes a comparison between LLMs and teaching assistants:

> *"But then again, I'm getting more comfortable with [ChatGPT]. So I'm using it more as a tool for understanding greater topics. […] Now it's I do a quick Google search. "That didn't come up with much. Let's see what ChatGPT says." So using it even more as a teaching assistant." [Par. 2]*

Participants were not polled during the interview whether this semester saw them interacting with course staff less than in the past, but these responses indicate that future interventions may benefit from that additional data point. While the idea that LLMs have the capacity to replace teaching assistants in certain contexts is present within some of the participants, whether or not they actively used LLMs in place of teaching assistants consistently has yet to be seen.

### *6. Over-reliance on LLMs*

Three students expressed concerns about potentially developing dependency on LLMs to complete their projects. In the following example, this student describes a scenario in where his dependence on LLMs might be detrimental:

*"Some days where I want to use it, I haven't been able to use it, which kind of made me realize it's easy to create a dependency on them and, if [online LLM services] go down, it's kind of annoying." [Par. 8]*

Par. 8 identifies that LLMs may be inaccessible due to external factors such as a web service failure. In this case, the student notes that being stripped of his ability to work with LLMs negatively affects his ability to complete his work. Another example suggests that excessive LLM usage may still cause concern even without external factors:

*"Yeah. I think things are a challenge for a reason, and you're supposed to be thinking and using your brain or else you get lazy. Over the summer, if you're just doing nothing, when you come back to school you're like, "I just forgot how to think." [Using LLMs is] like that." [Par. 4]*

Par. 4's suggestion that LLMs can reduce the difficulty of a challenge is not unique among other student responses. However, he warns that using LLMs too much can create a reliance that encourages behaviors in students that is detrimental to their own learning. For Par. 4 specifically, the idea that "forgetting" how to "think," or generally problem solve is enough to warrant his concerns about relying on LLMs too heavily.

One of the interview questions asked students to identify at what point in the ECE curriculum they felt LLMs should be formally introduced as a resource. Students' responses shed further light on the possibility of over-reliance on LLMs. Generally, responses fell into one of two camps: either that upper-level courses like ECE 461 should be where LLMs are introduced, or that lower-level classes that emphasize foundational skills are more appropriate. In this example, a student shares his insights:

*"I think this is a great class to have LLMs, because in the actual class, we're learning about the software development processes, cycles, tools, and techniques. We're not actually learning about the code or what we're actually making. So I think for a class like ECE 461, it is a great tool to use. However, for classes like [first- and second-year courses], I feel like those would not be as good because there's a template and it's really easy to copy and paste [solutions generated by the LLM]." [Par. 8]*

Here, Par. 8 compares the learning outcomes between upper- and lower-level courses. His mention of "templates" refers to the more uniform structure of assignments in introductory coding courses where we expect less variance in students' solutions when compared to project-based courses like ECE 461. While Par. 8 does not demonstrate any confidence in freshman and sophomore students to use LLMs responsibly, another student felt that introducing LLMs as early as possible is imperative:

*"But I know that if you introduce it too early, people are just going to use it to generate code, and, especially at the lower-level classes like data structures and algorithms, you can get by that class just by using LLMs to create code for you. Start [introducing LLMs] in freshman year, but have in-person exams where ChatGPT will not help them. [...] In [an] ideal world,*

> *students would learn about the responsibilities about using ChatGPT [from freshman classes]." [Par. 6]*

Par. 6 acknowledges that freshmen are capable of abusing LLMs in a way that Par. 8 warned about. However, Par. 6 understands the need to educate students about responsible LLM usage, and to take precautionary measures to prevent usage that would cause academic integrity problems.

## Discussion

In summary, our research aimed to explore how software engineering students interacted with LLM technologies during a semester-long project. Although each student we interviewed had unique experiences working with ChatGPT and/or Copilot, there were thematic similarities in many of their responses. Here we synthesize our observations and their implications.

First, as a research group, we underestimated the level of maturity and nuance with which undergraduate students would relate to LLMs. Interviewees were aware of the hazards of LLMs, such as the accuracy of generated content, the ethical dilemma of plagiarism, and the many societal, corporate, and individual perspectives on intellectual property theft. What tasks students deemed worthy of LLM aid varied between responses as well. Tasks that some students considered "grunt work" and felt comfortable assigning the LLM to solve, others felt was meaningful work and insisted on doing themselves.

In response to RQ1, we found that all interviewees used LLMs for coding support. More than half of the interviewees also used LLMs for writing support (e.g., emails and reports) or idea generation, both technical and conceptual. A significant minority of students also used LLMs as project managers. These findings are significant because although some previous studies have explored how students could use LLMs in their academic pursuits [12], this study is the first to observe how engineering students use LLMs in a context that both mandates but does not restrict their usage to particular use cases. Additionally, project management manifested as a unique use case from this study, which has not been observed in previous relevant literature. This discovery may be related to the large scope of the course project, as well as the lack of formal project management training in the curriculum.

In response to RQ2, we found that students hold a diversity of beliefs about how LLMs impact their learning. Students commonly believed that LLMs enhanced the pursuit of new knowledge, both in terms of information gathering efficiency and the implementation of that knowledge in their work. Equally as common was the concern among students that LLMs may negatively influence their ability to retain their newly discovered information. Making note that many participants held both beliefs simultaneously, we noticed that students shared a general sense of cautious optimism that the responsible use of LLMs can be a boon to learning, and the irresponsible use can have the opposite effect.

One group credited LLMs with improving accessibility of knowledge, another group credited LLMs with improving accessibility of solution, and yet another group expressed their concerns of knowledge retention when using LLMs. Two-thirds of the interviewees believed that gaining value from LLM usage is highly dependent on their relevant prerequisite knowledge. Just under one-half of the interviewees attributed their decreasing number of interactions with instructors and teaching assistants to the self-sufficient nature of LLMs. One-third of interviewees made mention of their concerns about becoming too reliant on LLMs to complete their engineering work. These findings reaffirm the background literature's mention of challenges related to LLM usage in academia like over-reliance and generated content bias [13]. However, they present new educational challenges not outlined in previous literature, namely the issue of feeling the need to have prerequisite knowledge before interacting with LLMs.

**Limitations**

The findings of the qualitative studies are driven by the context in which the study takes place. Therefore, the findings from a qualitative study like ours are not generalizable but instead may be transferable based on the context. Educators should take this into consideration when assessing the potential to employ a similar LLM policy in their classrooms. As our study was conducted with a specific group of upper-level male students in the ECE department at Purdue, the results of our study cannot be directly transferable or applicable to students that did not participate in the study or did not have similar backgrounds.

Another limitation of our study is the lack of gender and academic diversity. All the students in our study were male computer engineering students and hence, our findings are limited to perspectives of upper level male ECE students in a large midwestern research university. Having a diverse demographic distribution for our study will help capture variety of student perspectives while using LLM's. Further, we intend on re-examining our participant recruitment strategy to develop an approach that promotes more diversity. Finally, given that this is a qualitative study which requires researcher to immerse themselves during and after data collection and analysis, the researchers prior experience and knowledge about LLM's could have a small influence on the study. To counter this limitation, the researchers frequently took memos and notes of their ideas and perception during the data collection and analysis phases.

**Future Work**

Exploratory studies such as ours are intended to characterize a phenomenon to identify opportunities for further study. One opportunity is to study how LLM's influence the learning abilities of students who use them for their coursework. As our study was situated in an upper-level software engineering course, it helped us understand the perspectives that junior- and senior-level ECE students had on using LLMs. Although these students had prior experience in software development through the course prerequisites, our results highlight those students felt over-reliant on using LLM's in their coursework. A few students suggested that this over-reliance on LLM's could be detrimental as it can impact their learning abilities of required SWE skills if introduced too early in the curriculum. Therefore, in our future work we aim to investigate how usage of LLM's influence the learning of students earlier in the curriculum, at the freshman and sophomore levels. Students will use LLMs regardless of policy, so we would like to understand how LLMs can be safely integrated into lower-level courses without hurting the learning outcomes of students new to programming. For example, one of our research team members shared some of the interview data with their freshman-level class this semester, with the hopes that hearing accounts from senior students about the dangers of misusing LLMs. Finally, we believe many such similar investigations across different contexts (e.g., university type, class size, student demographic variations) are necessary to understand the right time to introduce LLM's in computer engineering curricula because LLMs when used effectively appear to have the potential to foster better learning outcomes.

Lastly, we see opportunity in studying student interactions with LLMs like ChatGPT as a source of feedback for a course. Prior studies have looked at the conversations that software engineers have with ChatGPT [42] and provided feedback on the behaviors and needs of practitioners. Similar studies can be conducted in educational contexts with students to uncover how student interactions with LLM's influence quality of students work product like assignments, projects, learning outcomes, etc. To effect this, a university would need a custom LLM (or a custom interface to a commercial LLM) that would anonymize, track, and summarize student queries to help instructors understand opportunities for improvement.

## Conclusion

In this paper, we reported the results of the first study on upper-level computing students using LLMs in a semester-long project. To conduct this exploratory study, we collected interview data on LLM-related experiences and perceptions from students enrolled in an upper-level software engineering course. We analyzed this interview data using thematic analysis approach to understand (1) how students used LLM applications when unrestricted by course policies and (2) uncover how students perceive the effects of LLM usage on their learning outcomes. Our findings reveal that students used LLMs to assist with technical (e.g., coding) and logistical (e.g., project management) aspects of their projects. Additionally, our study found that students perceived LLMs to greatly aid in their abilities to locate knowledge, create solutions, and work independently. However, students also reported to be concerned with developing an overdependence on LLMs thereby weakening their ability to retain knowledge. Our findings on the usage of LLMs in the software engineering landscape can help educators explore the role of this emerging technology in their respective academic settings.

## Acknowledgments

We thank S. Sinha for providing figures for ChatGPT and GitHub Copilot. We thank P. Jajal for his assistance in describing modern Large Language Models. We thank the students in ECE 461 for their cooperation. This work was funded by the Purdue Engineering Education Explorers Program, by an "AI in Teaching and Learning" grant from the Purdue Office of the Provost, and by a pedagogy grant from the Elmore Family School of Electrical and Computer Engineering.

# Appendix

## A1
*Round 1 interview protocol (~60 minutes).*

Introductory Remarks

This study is about the use of large language models (LLMs) in software engineering. That includes both "Q&A" LLMs such as ChatGPT, and tools that use LLMs as an underlying technology, such as GitHub Copilot. Please consider both kinds of LLMs in your answers.

Experiences pre-ECE 461

First, I will be going over a few questions on your experience using LLM tools prior to ECE 461.
1. How did you first find out about LLMs?
2. What made you want to explore LLM tools?
3. Can you describe your initial impressions of LLMs when you first started using them?
4. Did you find it easy or challenging to get started?
5. Before ECE 461, what experience did you have working with LLMs?
    a. In coursework at Purdue (permitted)
    b. In coursework at Purdue (not permitted)
    c. at internship(s)
    d. Hobby projects
6. Do you have any particular preferences with using LLM tools?
    a. (if yes to 6) How have your preferences evolved?
    b. (if yes to 6) What are the reasons for your LLM preference?
7. Did you receive any training or study any materials on how to use LLMs effectively in software engineering work? If yes, what did this training look like?

Experiences in ECE 461

Now, I will ask you a few questions on your experience using LLM tools in ECE 461.
1. What were your initial expectations or goals for using LLMs in this project?
2. What LLM do you intend to start this project with, and what motivates you to choose it for this project?
3. Are there any challenges or concerns you anticipate in using LLMs for this project?
4. Did you find the training and LLM homework provided by Prof. Davis to expand your understanding of the capabilities of LLMs?
5. As part of your team's plans for Phase 1 and Phase 2, your team had to describe their use of LLM technology.
    a. What was your team's policy?
    b. How did you decide on this policy?
6. Tell us about your typical use of LLMs in this project, and the kinds of value you get from them.

Interpretation of Experiences

Finally, I will ask you a few questions on how you interpret those 461 experiences.
1. Do you feel that the use of LLMs will enhance your learning experience? Why or why not?
2. How do you see the skills or knowledge you will gain from using LLMs benefiting your future coursework or projects?
3. At what point do you think it is appropriate to introduce LLMs into a software engineering curriculum?

**A2**
*Round 2 interview protocol (~60 minutes).*

Introductory Remarks

This study is about the use of large language models (LLMs) in software engineering. That includes both "Q&A" about LLMs such as ChatGPT, and tools that use LLMs as an underlying technology, such as GitHub Copilot. Please consider both kinds of LLMs in your answers.

Experiences in ECE 461

1. What are your current impressions of LLMs as compared to when you first started using them?
    a. How have your perceptions and impressions evolved?
    b. What changes do you notice between your previous and current usage of LLMs?
2. Have your LLM preferences over the course of the project changed between phase 1 and phase 2? Why?
3. How did you end up using LLMs differently throughout the course of the project?
    a. Was there a substantial difference in how they were used between phases?
4. How did you choose which tasks you would use LLMs to complete?
5. Are there any challenges or concerns you had while using LLMs for this project?
6. Have you had to seek help or support while using LLMs in your project? What was the nature of the issue or support, and how was it resolved?
7. How would you compare the use of LLMs in this project to its use in other academic tasks or projects?
8. Can you provide an example of a task or aspect of your project that was significantly improved by using LLMs?
9. Tell me about a time when you used LLMs in your project to accelerate your engineering work.
10. Did you experience a time when you used LLMs in your project and they slowed down your engineering work?
    a. Please describe any limitations you encountered in your experience.
11. Has using LLMs generally been an accelerator, a distraction, or …?
    a. Please describe any limitations you encountered in your experience.
12. How has using LLMs influenced the time required to complete this project?
13. How has using LLMs influenced the quality of your final project deliverables?

Interpretation of Experiences

Finally, I want to understand how you interpreted your experiences in 461.
1. Do you feel that the use of LLMs enhanced your learning experience? Why or why not?
2. How do you see the skills or knowledge you gained from using LLMs benefiting your future coursework or projects?
    a. How would LLM(s) be useful after graduation while working in an industry?
3. What alternative functionalities have you explored?
    a. Are there any features or functionalities of LLMs that you feel are underutilized or overlooked by students?
4. Looking back, would you make any changes in how you used LLMs for this project? If so, how would you approach them differently?
5. Is there anything else you would like to share about your experiences with the LLM in this project?